\documentstyle[12pt,twoside,fleqn,espcrc1,epsfig]{article}
\begin{document}
\input{epsf.tex}
\newcommand{\be}{\begin{eqnarray}}
\newcommand{\ee}{\end{eqnarray}}
\newcommand{\ben}{\begin{eqnarray*}}
\newcommand{\een}{\end{eqnarray*}}
\def\J{$J/\psi$}
\def\j{J/\psi}
\def\P{$\psi'$}
\def\p{\psi'}
\def\U{$\Upsilon$}
\def\u{\Upsilon}
\def\C{{\bar c}c}
\def\b{{\bar b}b}
\def\q{{\bar q}q}
\def\Q{{\bar Q}Q}
\def\F{$\Phi$}
\def\f{\Phi}
\def\T{$\hat t$}
\def\d{$d{\hat{\sigma}}$}
\def\t{\tau}
\def\E{\epsilon}
\def\a{\alpha_s}
\def\d{\lambda}
\def\la{\Lambda_{\rm QCD}}
\def\T{$T_f$}
\def\n{$n_b$}
\def\L{\cal L}
\def\l{\lambda}
\def\e{\epsilon}
\def\lsim{\raise0.3ex\hbox{$<$\kern-0.75em\raise-1.1ex\hbox{$\sim$}}}
\def\gsim{\raise0.3ex\hbox{$>$\kern-0.75em\raise-1.1ex\hbox{$\sim$}}}
%
\def\CMP{{ Comm.\ Math.\ Phys.\ }}
\def\NP{{ Nucl.\ Phys.\ }}
\def\PL{{ Phys.\ Lett.\ }}
\def\PR{{ Phys.\ Rev.\ }}
\def\PRep{{ Phys.\ Rep.\ }}
\def\PRL{{ Phys.\ Rev.\ Lett.\ }}
\def\RMP{{ Rev.\ Mod.\ Phys.\ }}
\def\ZP{{ Z.\ Phys.\ }}
\def\etal{{\sl et al.}}
\begin{flushright}
BI-TP 96/42\\
hep-ph/9609260
\end{flushright}
\vskip0.5cm
{\large Quarkonium Production and Colour Deconfinement in Nuclear 
Collisions}\footnote{Invited talk given at the ``Quark Matter 96" Conference, 
 Heidelberg, May 20-24, 1996}
\vskip0.8cm
\begin{flushleft}
D. Kharzeev\footnote{Work supported by GSI under project BISAT.}\\
\vskip0.3cm
Fakult\"at f\"ur Physik, Universit\"at Bielefeld, D-33501 Bielefeld, Germany
\end{flushleft}
\vskip0.3cm
\begin{abstract}
The suppression of quarkonium production in nucleus-nucleus collisions was 
originally proposed as a signal of 
colour deconfinement. Strong ``anomalous" \J~ suppression in Pb-Pb collisions 
has been reported at this Conference by the NA50 Collaboration. 
Is this suppression really anomalous? 
Can we conclude that the quark-gluon plasma is already discovered? 
What has to be done next? I address these questions basing on the current 
theoretical understanding of quarkonium production and new 
precise experimental information. 
\end{abstract}

\section{INTRODUCTION}

The idea to use heavy quarkonia as a probe of 
excited QCD matter produced in relativistic heavy ion 
collisions was proposed a decade ago \cite{MS}. 
This suggestion was based on the concept 
of colour screening of static potential acting between the heavy quark 
and antiquark, which occurs in hot and/or dense quark-gluon matter. 
A year later, the \J\ suppression in nucleus-nucleus collisions was observed 
experimentally \cite{NA38}.
Since that time, quarkonium suppression has always been a respectable 
prospective signal 
of deconfinement. However, the experimentally observed suppression has been 
consistent not only with the deconfinement scenario, but with  
a plenty of ``conventional" explanations as well \cite{Blaizot}.   
Moreover, the differences between various conventional approaches, 
invoking such different mechanisms as nuclear absorption, 
gluon shadowing, or comover absorption, but nevertheless all providing 
more or less reasonable description of the data, created a lively 
controversy.  
The problem thus was clearly awaiting a more detailed and systematic 
analysis, many essential inputs for which were missing. 
In fact, the lack of precision data on \J\ and \P\ production 
in $p-A$ collisions, poor theoretical understanding of \J\ production and 
its interactions with light hadrons made this analysis 
virtually impossible.

Fortunately, the situation has started to change recently: 
new high precision data 
on quarkonium production (see \cite{Carlos}-\cite{San} and 
references therein) have significantly advanced the theory \cite{Braa}, 
and the operator product expansion techniques  
have allowed a systematic 
calculation of quarkonium absorption cross sections \cite{Peskin}-\cite{KS} 
and dissociation rates in confined matter \cite{KS}-\cite{K1}. 
The dominance of higher Fock states in quarkonium production, revealed 
by the new Fermilab data \cite{San} and naturally 
emerging theoretically, has inspired a new 
approach to nuclear attenuation of quarkonium production \cite{KS2}. 
This approach, 
as I shall discuss in this talk, has 
enabled a quantitative understanding of quarkonium suppression in 
both $p-A$ and $A-B$ collisions, giving a credit to the nuclear absorption 
model \cite{GH}. It has become clear that the existing 
$S-U$ data in fact do {\it not} exhibit any anomalous behaviour 
and are {\it not} consistent with a deconfinement scenario, which  
requires additional strong suppression. This is why the results  
on \J~ suppression in Pb-Pb collisions were so anxiously awaited -- 
they were our last chance to see something unusual before the advent of 
future experiments at RHIC and the LHC.

These results, presented by the NA50 Collaboration at this 
Conference \cite{MG}, are striking. 
The data clearly show a strong \J~ suppression, 
going way beyond the expected. But is this suppression really anomalous? 
Have we finally reached the border of the long-awaited {\it terra incognita} 
of deconfined quark-gluon matter? In this talk, I shall attempt to 
address these questions.       

\section{QCD ATOMS IN EXTERNAL FIELDS}

\subsection{Quarkonium Interactions and the Operator Product Expansion}  

\vskip0.3cm
In the Operator Product Expansion (OPE) approach, 
the amplitude of heavy quarkonium
interaction with light hadrons is represented in the form
\be
F_{\f h} = i\int d^4x e^{iqx} \langle h|T\{J(x)J(0)\}|h \rangle =
\sum_n c_n(Q,m_Q) \langle O_n \rangle ,
\label{2.5}
\ee
where the set $\{O_n\}$ should include all local gauge-invariant operators
expressible in terms of gluon fields; the matrix elements $\langle
O_n \rangle$ are taken between the initial and final light-hadron
states. The coefficients $c_n$ are expected to be computable perturbatively
and are process-independent. 
\vskip0.3cm
\begin{minipage}[t]{6cm}
\setlength{\unitlength}{0.00800in}%
\begingroup\makeatletter\ifx\SetFigFont\undefined%
\gdef\SetFigFont#1#2#3#4#5{%
  \reset@font\fontsize{#1}{#2pt}%
  \fontfamily{#3}\fontseries{#4}\fontshape{#5}%
  \selectfont}%
\fi\endgroup%
\begin{picture}(229,207)(83,535)
\thinlines
\put(270,715){\oval(20,50)}
\put(200,570){\oval(110,70)}
\put( 85,740){\line( 1, 0){225}}
\put( 85,690){\line( 1, 0){225}}
\multiput(160,690)(0.00000,-8.26087){12}{\line( 0,-1){  4.130}}
\put(130,715){\oval(20,50)}
\multiput(240,740)(0.00000,-7.83784){19}{\line( 0,-1){  3.919}}
\put(255,560){\line( 1, 0){ 55}}
\put( 85,580){\line( 1, 0){ 60}}
\put( 85,570){\line( 1, 0){ 60}}
\put( 85,560){\line( 1, 0){ 60}}
\put(255,580){\line( 1, 0){ 55}}
\put(255,570){\line( 1, 0){ 55}}
\end{picture}
\end{minipage}
\begin{minipage}[b]{9cm}
Figure 1. A sample diagram describing quarkonium 
interaction with a 
light hadron in the OPE scheme; dashed lines are the gluon propagators, 
ovals represent the quarkonium wave function, and the blob stands for  
the gluon structure function of the hadron.
\end{minipage} 
\vskip0.3cm
 The Wilson coefficients $c_n$ were computed for S \cite{Peskin} and 
P \cite{K1}, \cite{KS4} states in the leading order in $1/N^2$ ($N$ is the 
number of colours).
The expectation values $\langle O_n \rangle$ of the operators composed
of gluon fields can be expressed as Mellin
transforms \cite{Parisi} of the gluon structure function of the light
hadron, evaluated at the scale $Q^2=\e_0^2$,
\be
\langle O_n \rangle = \int_0^1 dx\ x^{n-2} g(x, Q^2 = \e_0^2).
\label{2.12}
\ee   
\begin{figure}[h]
\begin{center}
\epsfxsize14cm
\mbox{\epsfig{file=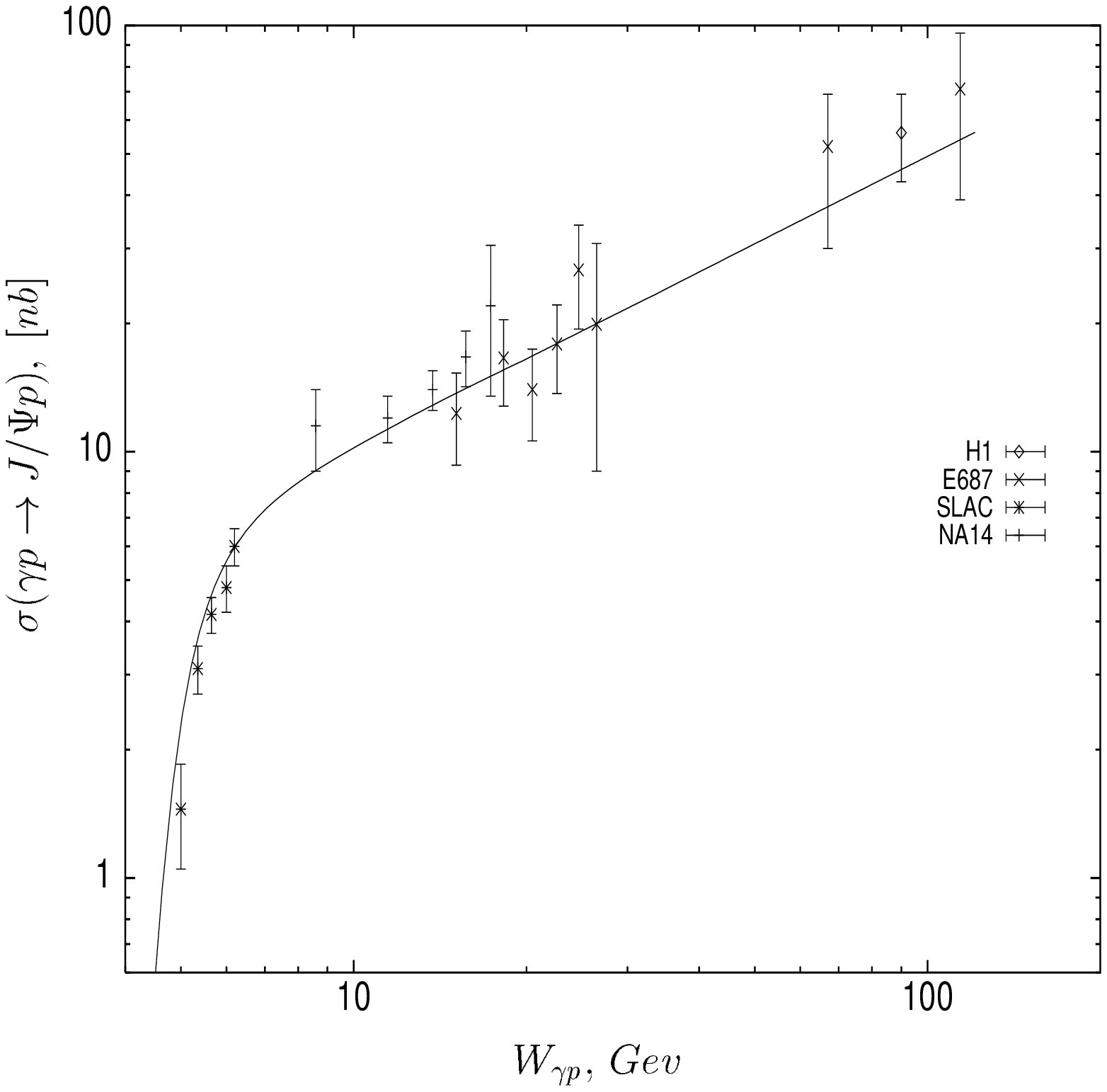,bbllx=2.cm,bblly=10.cm,bburx=21.5cm,bbury=26.cm,width=14.cm,angle=0}}
       \end{center}
\noindent Figure 2. $J/\psi$ photoproduction cross section; 
the curve is the theoretical prediction \cite{KSSZ}.
\label{prod}
\end{figure}

Since the total $\f-h$ cross section is proportional
to the imaginary part of the amplitude $F_{\f h}$, the dispersion
integral over the c.m.s. energy $\lambda$ leads to the set of sum rules, 
relating the cross section to 
the gluon structure function of the light hadron. 
This relation, illustrated in Fig.\ 1, has 
 a very important property: the magnitude and energy 
dependence of the quarkonium dissociation 
cross section at low energies is entirely determined by the behaviour 
of the gluon structure function at large $x \sim 1/\lambda$, 
whereas the cross section at high energy is governed by the small $x$ 
behaviour of the structure function. Since the gluon structure functions 
of light hadrons are suppressed at large $x$, the calculated cross section 
rises very slowly from the threshold. When the 
hadron momentum in the $J/\psi$ rest frame is $P_h \simeq 5 $ GeV, 
the cross section is more than an order of magnitude 
below its asymptotic value.

Recently, the calculation sketched above has been refined \cite{KSSZ} 
by taking 
into account target mass corrections, the real part of the scattering
amplitude restored by dispersion relations, and the use of modern gluon 
structure functions inferred from the analyses of HERA data. 
This allows to evaluate the cross section in the entire energy range 
accessible to present experiments;
the results confirm the threshold behaviour of the absorption cross 
section established previously. 
Vector meson dominance relates the cross sections of 
\J~ dissociation and photo-production; Fig.\ 2 shows the results compared 
to the available data. One can see that a strong threshold suppression 
of the \J~ absorption cross section is actually required by 
the data. 

\subsection{Quarkonium Interactions with Pions}

\vskip0.3cm
It can be shown that spontaneously broken chiral and 
scale symmetries of QCD imply decoupling of low-energy pions from heavy 
quarkonium \cite{K1}.  
The proof is based on the application of low-energy 
QCD theorems \cite{JE} (see \cite{EK} for a recent review and 
introduction) to the amplitude of quarkonium interactions with 
light hadrons \cite{VZ}. 
Qualitatively, the origin of the decoupling can be explained in the 
following way. 
At low energies, the amplitude of quarkonium interaction is proportional to 
the gluon field operator dominating the trace of the energy-momentum tensor of 
QCD.
The appearance of this operator in the trace of the energy--momentum 
tensor is a reflection of 
the broken scale invariance of QCD, so the coupling is determined by 
the scale dimension of the hadron field. Chiral symmetry, however, implies 
zero scale dimension for the Goldstone boson fields -- 
otherwise scale 
transformations would break chiral invariance. 

\subsection{Quarkonium Production in Hadron Collisions}
\vskip0.3cm
The perturbative approach to quarkonium production \cite{CS},\cite{Ger} 
is based on the assumption 
that the production process is localized at distances $\sim m_{Q}^{-1}$, 
much shorter than the size of quarkonium $r \sim [\alpha_s(r^{-1})m_{Q}]^{-1}$.
This approach is justified if all 
gluons involved in the production carry a high momentum 
$q \sim m_{Q}$. However, the hadroproduction of vector states, 
for example, requires at least three gluons, of which only two must be hard 
to create the $\bar{Q}Q$ pair. At small $P_T$ (the domain that dominates 
the integrated quarkonium production cross sections), the third gluon 
can be very soft, and is emitted (or absorbed) 
at distances of the order of quarkonium size. This is 
clearly inconsistent with the factorized form of the amplitude, and may 
``explain" the failure of perturbative approach in describing 
the integrated cross sections of 
quarkonium production at fixed target energies. 
At collider energies, the perturbative approach fails even at high $P_T$, 
since the non-perturbative contribution to the gluon fragmentation 
becomes important \cite{Braa}. These arguments help to understand 
the phenomenological success of the colour 
evaporation model in explaining the data (see \cite{Gav} for a recent study).

A consistent solution of this problem emerges if one assigns the soft gluon to 
the quarkonium wave function introducing the notion of 
$|\bar{Q}Qg...\rangle$ higher Fock states \cite{Braa}. 
 In fact, such states appear naturally in the OPE scheme 
described above. Consider, for example, the amplitude of quarkonium 
interaction with an external gluon field (see Figs.\ 1 and 3): 
it includes the transformation of the colour-singlet quarkonium 
into a colour-octet $[\bar{Q}Q]_8$ 
state. The overall colour neutrality is of course preserved and ensured 
by the coloured gluon cloud surrounding the $[\bar{Q}Q]_8$ state. 
The simplest example of such system is provided by the 
$|[\bar{Q}Q]_8 g\rangle$ 
state. Since the vacuum of QCD has a complicated structure \cite{SVZ} 
with $\langle 
g^2 G^2 \rangle \neq 0$, it induces a significant admixture of the 
$|[\bar{Q}Q]_8 g\rangle$ component in the wave function of quarkonium 
\cite{Vol} -- see Fig. 3a.
\begin{figure}
\vskip0.5cm
\begin{center}
\setlength{\unitlength}{0.008in}%
\begingroup\makeatletter\ifx\SetFigFont\undefined%
\gdef\SetFigFont#1#2#3#4#5{%
  \reset@font\fontsize{#1}{#2pt}%
  \fontfamily{#3}\fontseries{#4}\fontshape{#5}%
  \selectfont}%
\fi\endgroup%
\begin{picture}(589,149)(83,593)
\thinlines
\put(270,715){\oval(20,50)}
\put(465,715){\oval(20,50)}
\put(635,715){\oval(20,50)}
\put( 85,740){\line( 1, 0){225}}
\put( 85,690){\line( 1, 0){225}}
\multiput(160,690)(0.00000,-8.26087){12}{\line( 0,-1){  4.130}}
\multiput(230,740)(0.00000,-7.83784){19}{\line( 0,-1){  3.919}}
\multiput(155,605)(0.40000,-0.40000){26}{\makebox(0.1111,0.7778){\SetFigFont{5}{6}{\rmdefault}{\mddefault}{\updefault}.}}
\multiput(155,595)(0.40000,0.40000){26}{\makebox(0.1111,0.7778){\SetFigFont{5}{6}{\rmdefault}{\mddefault}{\updefault}.}}
\multiput(225,605)(0.40000,-0.40000){26}{\makebox(0.1111,0.7778){\SetFigFont{5}{6}{\rmdefault}{\mddefault}{\updefault}.}}
\multiput(225,595)(0.40000,0.40000){26}{\makebox(0.1111,0.7778){\SetFigFont{5}{6}{\rmdefault}{\mddefault}{\updefault}.}}
\put(425,740){\line( 1, 0){ 90}}
\put(130,715){\oval(20,50)}
\put(515,740){\line( 0,-1){ 50}}
\multiput(610,605)(0.40000,-0.40000){26}{\makebox(0.1111,0.7778){\SetFigFont{5}{6}{\rmdefault}{\mddefault}{\updefault}.}}
\put(425,690){\line( 1, 0){ 90}}
\put(585,740){\line( 0,-1){ 50}}
\put(585,740){\line( 1, 0){ 85}}
\put(585,690){\line( 1, 0){ 85}}
\multiput(515,740)(8.23529,0.00000){9}{\line( 1, 0){  4.118}}
\multiput(515,690)(8.23529,0.00000){9}{\line( 1, 0){  4.118}}
\multiput(485,690)(0.00000,-8.26087){12}{\line( 0,-1){  4.130}}
\multiput(615,690)(0.00000,-8.26087){12}{\line( 0,-1){  4.130}}
\multiput(480,595)(0.40000,0.40000){26}{\makebox(0.1111,0.7778){\SetFigFont{5}{6}{\rmdefault}{\mddefault}{\updefault}.}}
\multiput(610,595)(0.40000,0.40000){26}{\makebox(0.1111,0.7778){\SetFigFont{5}{6}{\rmdefault}{\mddefault}{\updefault}.}}
\multiput(480,605)(0.40000,-0.40000){26}{\makebox(0.1111,0.7778){\SetFigFont{5}{6}{\rmdefault}{\mddefault}{\updefault}.}}
\end{picture}
\end{center}
\hskip6.5cm a) \hskip6.8cm b)
\vskip0.3cm
\noindent Figure 3. Interactions of quarkonium with external 
gluon fields; solid (dashed) lines are the heavy quark (gluon) 
propagators, and ovals represent the quarkonium wave function.
\end{figure} 
For a physical $J/\psi$ state, this leads to the following generic 
decomposition:
\be
|J/\psi\rangle = a_1\ |\bar{c}c\rangle\ + \ a_2\ |[\bar{c}c]_8 g\rangle \ +\ ... \label{fock}
\ee
Similar decompositions hold for other quarkonium states; for $\chi$ states, 
for instance, the importance of higher Fock component is implied by 
the divergence of the perturbative annihilation amplitude in the soft 
gluon limit \cite{NR}.  
The magnitude of the $|[\bar{Q}Q]_8 g\rangle$ state 
 admixture is reflected  by the 
magnitude of relativistic corrections in the NRQCD approach \cite{NR} and 
by the size of power corrections in the QCD sum rule approach \cite{SVZ}. 
These corrections are generally not very large, making 
applicable the familiar  
concept of heavy quarkonium as of a non-relativistic system essentially 
composed of just $\bar{Q}Q$ state. However in certain processes -- like 
production and annihilation of quarkonium -- these components can 
play extremely important role\footnote{Another example is provided by 
the scattering of quarkonium states at very high energies \cite{Mue}.}. 
In fact, the leading order production of heavy 
vector quarkonium  
proceeds via the gluon fusion producing the $\bar{Q}Q$ pair in a 
colour-octet state that later neutralizes its colour emitting (or absorbing) 
an extra gluon. If this extra gluon is soft (as is the case in the small 
$P_T$ domain), the production process can be visualized as proceeding 
via the higher Fock state $|[\bar{Q}Q]_8 g\rangle$ (see Fig.3b).  
 
Since the colour Coulomb 
interaction between the heavy quarks in the colour-octet state is repulsive 
and weak ($\sim 1/(N^2-1)$ 
with respect to the attraction in the colour-singlet 
state, where $N$ is the number of colours), the $|[\bar{Q}Q]_8 g\rangle$ 
state is separated from the basic 
$|\bar{Q}Q\rangle$ state by the mass gap of $\simeq \epsilon_0$, where 
$\epsilon_0$ is quarkonium binding energy. This (virtual) state therefore 
has a proper lifetime of $\tau \simeq 1/\epsilon_0$.  
In the frame where quarkonium moves with momentum $P$, the superposition  
(\ref{fock}) will be coherent over a distance $z_c \simeq \tau P/2M_{Q}$. 
At high energies, this distance is sufficient for a produced 
$|[\bar{Q}Q]_8 g\rangle$ state to traverse the entire nuclear volume. 

What will be the effect of the nuclear medium on the propagation of 
such a state? 
To answer this question, let us first note that  
the produced $[\bar{Q}Q]_8$ pair is initially almost pointlike, with the 
transverse size of $r^{\bar{Q}Q}_{\perp}\sim 1/2m_Q$ (see Fig. 3b). 
The produced $|[\bar{Q}Q]_8 g\rangle$ state can be thus considered as 
a colour dipole formed by an almost pointlike colour-octet 
$\bar{Q}Q$ state and 
a collinear gluon. 
The transverse size of the $|\bar{c}c g\rangle$ state can be estimated 
\cite{KS2} from 
the characteristic virtualities of the diagram of Fig.3b as 
$r_{\perp} \simeq (2m_c \Lambda_{{\rm QCD}})^{-1/2} \simeq 0.20-0.25$ fm. 
An interaction inside nuclear matter 
will most likely prevent this state from binding, at later stage, to 
the quarkonium -- the colour octet $\bar{c}c$, with its collinear gluon 
stripped off, 
will preferably produce open charm mesons\footnote{Note that 
the \J~ production cross section represents only a tiny 
part, of the order of 1\%, of the total charm production -- this means that 
the probability to pick up a collinear gluon for the colour-octet $\bar{Q}Q$ 
state is in general very small.}.

Let us try to estimate the break-up cross section of such 
$|[\bar{Q}Q]_8 g\rangle$ state in its interaction with nucleons \cite{KS2}. 
The 
transverse size of the $|[\bar{c}c]_8 g\rangle$ state estimated 
above is roughly the same as the size 
of $J/\psi$. The $|[\bar{c}c]_8 g\rangle$ state however is not bound, so, 
contrary to the case of $J/\psi$, we do not expect any threshold suppression 
of the break-up cross section. 
We can therefore estimate the $|[\bar{c}c]_8 g\rangle$ break-up 
cross section rescaling the value of the $J/\psi$ break-up cross 
section at high energy (where the 
threshold suppression does not affect the cross section) 
by the colour factor $9/4$, 
arising from the difference between the couplings of colour dipoles 
formed by the triplet and octet charges. At the energy range relevant for 
the fixed target experiments, the $J/\psi$ break-up cross section evaluated 
in the formalism of section 2.1 is $\sigma_{J/\psi N} \simeq 2.5-3$ mb. 
We therefore get 
$\sigma_{(\bar{c}c g) N} \simeq 6 - 7$ mb as an estimate of the 
$|[\bar{c}c]_8 g\rangle$ absorption cross section. The analogous estimate for 
bottomonium states yields $\sigma_{(\bar{b}b g) N} \simeq 1.5 - 2$ mb.      
These estimates are admittedly rough; they show, however, that the 
nuclear attenuation of quarkonium production is in general quite strong and, 
in the first approximation, is universal for various quarkonium states.

\section{QUARKONIUM AS A PROBE OF DECONFINED MATTER}

We have shown in the previous section that the absorption cross sections 
of tightly bound quarkonium states at low energies are very small 
due to the softness of gluon fields confined inside light 
hadrons; 
this protects \J~ in a thermal hadron gas at all meaningful 
temperatures ($T\leq 0.3$ GeV) \cite{KS}-\cite{K1}.
On the other hand,  
the distribution of gluons in a deconfined medium is
directly thermal, so that the deconfined gluons are hard, with the average 
momentum of $\langle p_g \rangle_{\rm deconf} = 3T$.
An immediate consequence of deconfinement is thus
a considerable hardening of the gluon momentum distribution \cite{KS,KS1}. 
Hard deconfined gluons can easily break up the \J; the cross section of this 
``gluo-effect" is given by
\be
\sigma_{g J/\psi}(k) = {2\pi \over 3} \left( {32 \over 3} \right)^2 \left(
{m_c \over \e_0} \right)^{1/2} {1 \over m^2_c} {(k/\e_0 - 1)^{3/2}
\over (k/\e_0)^5 }, \label{5.4}
\ee
where $k$ is the momentum of the gluon incident on a stationary
quarkonium with binding energy $\epsilon_0$.
We thus see qualitatively how a deconfinement test can be carried out.
If we put a \J~into matter at a temperature $T=0.2$ GeV, then
the \J~will survive if the matter is confined, and will disappear if 
the matter is deconfined, since in the latter case the gluons will be 
hard enough to break it up.

The latter part of this statement is in accordance with the original prediction
that the formation of a QGP should lead to a \J~suppression
\cite{MS,KMeS}. There it was argued that in a QGP, colour screening
would prevent any
resonance binding between the perturbatively produced $c$ and ${\bar
c}$, allowing the heavy quarks to separate. At the hadronization point
of the medium, they would then be too far apart to bind to a \J~and
would therefore form a $D$ and a $\bar D$. Our picture complements this 
argument by the conclusion that additional suppression 
of physical  \J\ 
in dense matter will occur {\it if and only if} there is deconfinement. 

The dissociation of \J\ (or $\Upsilon$) in both pictures is a 
consequence of the interaction with strong 
gluon fields present in deconfined matter. There is, however, 
a difference between the two mechanisms: the static screening picture 
takes into account the effect of deconfined fields on the binding potential 
acting between the heavy quarks, but neglects the energy-momentum transfer 
between the \J\ and the heat bath. The dynamical ``gluo-effect" picture 
of quarkonium suppression, 
on the other hand, emphasizes the role of the energy-momentum transfer 
from deconfined gluons to the \J~, but neglects the screening 
of the binding potential. Both pictures are expected to describe the 
physics of \J\ suppression in their respective domains of applicability; 
they should emerge as two limits in one unified microscopic approach, 
that still has to be developed. The parameter that is relevant in this problem 
is $X(T) \equiv \Delta E(T) / T$,
where the binding energy of quarkonium $\Delta E$ depends on the temperature 
of the system $T$ because of the Debye screening. In the weak coupling 
limit of $X\ll 1$, 
the binding energy is negligible compared to the temperature, and 
the quarkonium will simply fall apart with the rate
$R = 4 / L (T / \pi M_Q)^{1/2}$
($L$ is quarkonium size), which is the classical high temperature 
limit of thermal activation rate \cite{KMS}. 
In the strong coupling limit of $X\gg 1$, 
on the other hand, the system is tightly bound, and the binding energy 
threshold has to be overcome by the absorption of hard gluons from 
the heat bath. 
The rate of dissociation in this case should be computed from the 
thermal average of the  gluon-quarkonium cross section (\ref{5.4}).
The actual value of $X$ at different temperatures depends, of course, 
on the detailed dynamics of screening; lattice calculations can be 
of significant help here, fixing the temperature dependence 
of quarkonium mass.

It is important to note that the dynamical ``gluo-effect" 
approach to \J~suppression does not
require a thermal equilibrium of the gluon fields, so that it
will remain applicable even in deconfined pre-equilibrium stages. 
Quarkonium 
interactions in an equilibrating parton gas were considered in ref. \cite{XW}.
 
\section{PHENOMENOLOGY OF QUARKONIUM PRODUCTION\\ 
IN NUCLEAR COLLISIONS$^4$}
\addtocounter{footnote}{1}\footnotetext{This section is 
based on the work \cite{KLNS}.} 
\subsection{$p-A$ collisions}
\vskip0.3cm
According to our discussion in section 2.3, 
in the presently accessible kinematic
region of \J~production by $p-A$ collisions ($x_F\geq 0$), the target
nucleus sees only the passage of the pre-resonance state; physical
charmonium states are formed outside the nucleus. 
The size of the pre-resonance state is determined by the charmed 
quark mass and confinement scale and is therefore the same for \J~and \P. 
The nuclear attenuation of \J~ and \P~ production in $p-A$ collisions  
should thus be universal. 
Indeed, the \J~and  \P~production in $pA$
collisions shows to the same $A$-dependence. Fitting the available data on 
the \P/(\J) ratio \cite{Carlos} to the form $A^{\alpha}$ leads to
\be
\alpha = 0.0 \pm 0.02, \ \ 95\%\ C.L.;\label{5}
\ee
this rules out variations of more than 10 \% between $pp$ and $pU$ collisions.
The suppression of \J~production in $p-A$ collisions should thus be
understood as pre-resonance absorption in normal nuclear matter.  
This accounts naturally for the
equal suppression observed for the two states, which would be impossible
for physical resonances of such different sizes.  

We shall now determine the pre-resonance absorption  
cross section from the NA38/51 $p-A$ data at incident
proton beam energies of 200 and 450 GeV \cite{Carlos}.
In Glauber theory, the survival probability for a \J~produced in a
$p-A$ collision is given by
\be
S_{pA}^{Gl} = {\sigma_{pA\to \psi}\over A \sigma_{pN \to \psi}} = \int
d^2b~dz
\rho_A(b,z) \exp\left\{-(A-1)\int_z^{\infty} dz' \rho_A(b,z')
\sigma_{abs} \right\}. \label{2}
\ee
Here $\rho_A$ is the nuclear density distribution, for which we take the
standard three-parameter Woods-Saxon form with parameters as 
tabulated in Ref.\ \cite{deJager};
it is normalized to unity, with $\int d^2b dz \rho_A(b,z) = 1$. 
The suppression is thus fully determined by the absorption
cross section $\sigma_{abs}$ in nuclear matter. From the NA38/51 data 
we obtain the best fit for
\be
\sigma_{abs} = 6.3 \pm 0.6~ {\rm mb}, \ \ 95\%\ C.L.; \label{4}
\ee
the corresponding
survival probabilities are plotted in Fig. 4. The agreement is
seen to be excellent in all cases. 
The value (\ref{4}) is consistent with the theoretical estimates of 
section 2.3, which suggest for the absorption cross section of the 
$\C-g$ on nucleons $\sigma_{abs} \simeq 6 - 7$ mb \cite{KS2}.
\begin{figure}[h]
\hskip1.4cm {\large{$S_{pA}^{J/\psi}$}}
\begin{center}
\epsfxsize12cm
\mbox{\epsfig{file=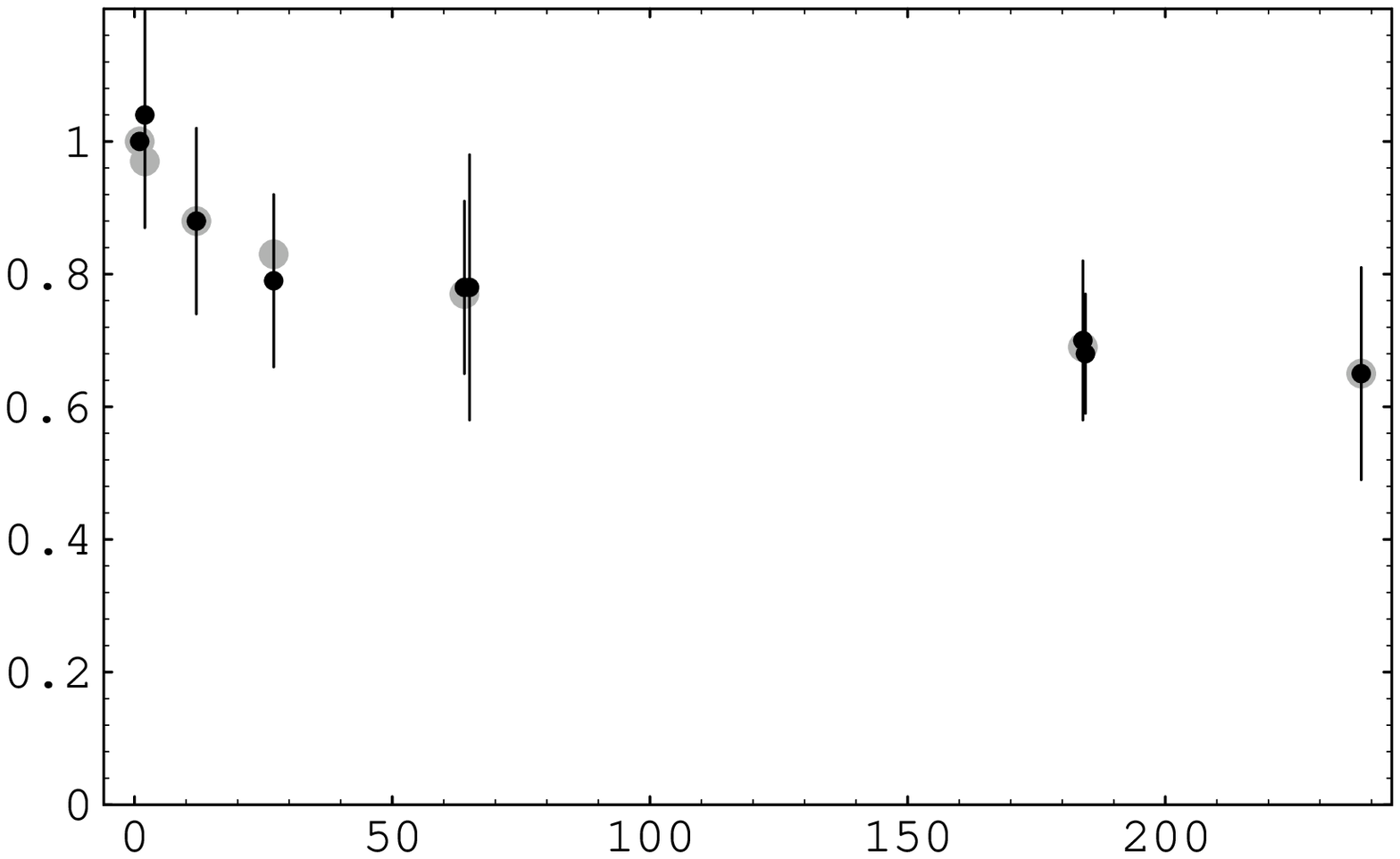,bbllx=1.cm,bblly=9.2cm,bburx=21.5cm,bbury=16.5cm,width=12.cm,angle=0}}
       \end{center}
\hskip12cm {\large{A}}\\
\noindent Figure 4. $J/\psi$ suppression in $pA$ collisions; the NA38/50 
data (black points) are compared to the Glauber theory calculations 
(grey points) with $\sigma_{abs}=6.3 \pm 0.6$ mb.
\end{figure}

We have carried out the same analysis for the 800 GeV 
E772 data (see \cite{Pat} for a review); here
the cross section is slightly larger: $\sigma_{abs} = 7.4 \pm 0.7$ mb , 
but within errors compatible with the value (\ref{4}) 
obtained from the NA38/51 data. 
A slow increase 
of the absorption cross section with energy can be attributed to the 
growth of the gluon structure function towards smaller $x$ 
(the same effect is responsible for the increase of the $J/\psi$ absorption 
cross section in the relevant energy range, see Fig.\ 2). 

We thus conclude that \J~and \P~production in $pA$ collisions is
quantitatively well described by absorption of a pre-resonance
charmonium state in nuclear matter, with the absorption cross section
for both states in the energy range of SPS experiments given by 
Eq.\ (\ref{4}). We now extend this description to nuclear collisions.

\subsection{{\it{\bf S-U}} Collisions}
\vskip0.3cm

In nucleus-nucleus collisions, charmonium production can be measured
as function of the centrality of the collision, and hence we have to
calculate the \J~survival probability at fixed impact parameter $b$. It
is given by 
\be
{ d S_{AB}^{Gl}(b) \over d^2 b} = 
{1 \over A B~ \sigma_{NN \to \psi}}\left[{d \sigma_{AB
\to \psi} \over d^2 b} \right] =  
 \int d^2s dz dz' \rho_A(\vec{s},z) \rho_B(\vec{b}-\vec{s},z') 
S_A(z,\vec{s}) S_B(z',\vec{s}), \label{6}
\ee
where $S_A(z,\vec{s}) = 
\exp\left\{-(A-1)\int_z^{\infty} dz_A~ \rho_A(\vec{s},z_A)~ \sigma_{abs}
\right\}$, and analogously for $S_B(z',\vec{s})$.
Here $\vec{s}$ specifies the position of the production point in a plane
orthogonal to the collision axis, while $z$ and $z'$ give the position
of this point within nucleus $A$ and within nucleus $B$, respectively.
The nuclear density distributions $\rho_A$ and $\rho_B$ are defined as
above. To obtain normalized survival probability at fixed impact parameter 
$b$, we have to divide $[dS^{Gl}/d^2b]$ by 
$[dS^{Gl}(b;\sigma_{abs}=0)/d^2b]$.   
\par
Experimentally, the centrality of the collision is determined by a
calorimetric measurement of the associated transverse energy $E_T$; 
 we thus have
to establish and test a correpondence between impact parameter $b$ and
transverse energy $E_T$.
This correlation can be expressed in terms of the number of ``wounded" nucleons
\cite{Bialas}. Each wounded nucleon contributes on the average an amount 
$q$ to the overall transverse energy produced in the collision, so we  
have the relation
\be
{\bar E}_T(b)~ =~q~{\bar N}_w(b) \label{propw}
\ee
between the average number ${\bar N}_w$ of nucleons wounded in a
collision at fixed impact parameter $b$ and the associated average
transverse energy ${\bar E}_T$ produced in that collision. In the
analysis of specific experimental results, the proportionality factor
$q$ depends on the details of the detector, in particular on the
rapidity and transverse momentum range in which the produced
secondaries are measured.
\par
The average number of wounded nucleons in an $AB$ collision at impact
parameter $b$ is given by 
\be
{\bar N}^w_{AB}(b)~\equiv \int d^2 s\ n^w_{AB}(b,s) = 
~& A \int d^2 s~ T_A(\vec{s}) \left\{ 1-[1-\sigma_N
T_B(\vec{s}-\vec{b})]^B\right\} + & \nonumber \\
+~ & B \int d^2 s~ T_B(\vec{s}-\vec{b}) \left\{ 1-[1-\sigma_N
T_A(\vec{s})]^A\right\}. & \label{9}
\ee
Here $\sigma_N\simeq 30$ mb denotes the inelastic 
production cross section, and
$T_A(\vec{s}) =\int dz~\rho_A(z,\vec{s})$ the nuclear profile 
function; the
$\vec{s}$-integration runs again over a plane orthogonal to the collision
axis. The distribution (\ref{9}) is normalized in the following way:
\be
{\bar N}^w_{AB} = {1 \over \sigma_{AB}} \int d^2 b {\bar N}^w_{AB}(b) = 
 {1 \over \sigma_{AB}}\ (A \sigma_B + B \sigma_A).
\ee

Since there are fluctuations in the number of wounded nucleons and in the 
transverse energy of the secondaries that each wounded nucleon produces,
there will be corresponding fluctuations in the relation between $E_T$
and $b$. We assume the dispersion $D$ in the produced transeverse energy 
to be proportional to $\sqrt{N_w}$, $D^2~ =~ a {\bar E}_T(b)$, 
with a universal physical parameter $a$ to be determined from
$pA$ or $AB$ collisions.  
We choose the $E_T-b$ correlation function $P_{AB}(E_T,b)$ as a 
conventional (see, e.g., \cite{GV}) Gaussian 
distribution around the central value (\ref{propw}) with dispersion $D$; 
it is normalized at fixed $b$: $\int dE_T ~P_{AB}(E_T,b) = 1$.  

We have checked \cite{KLNS} that both minimum bias \cite{Stock}  
and Drell-Yan associated \cite{NA38},\cite{MG} transverse energy 
spectra are very well reproduced in the approach outlined above.
With the relation between the measured transverse energy $E_T$ and the
impact parameter $b$ of the collision thus determined, we can now
calculate the $E_T$ dependence of the charmonium survival probability
in nuclear matter. 

We begin with \J~production. The experimentally determined quantity 
is the ratio $(d\sigma_{AB}^{J/\psi}/dE_T)/(d\sigma_{AB}^{DY})$ of $J/\psi$ 
to Drell-Yan production, measured in the mass interval 
$2.9 \leq M_{\mu\mu} \leq 5.5$ GeV. From this we obtain the survival 
probability at fixed $E_T$
\be
S^{J/\psi}_{exp}(E_T) = {\sigma^{DY}_{AB} \over \sigma^{J/\psi}_{AB}}\ 
\left[{ d\sigma_{AB}^{J/\psi} \over dE_T} / 
{ d\sigma_{AB}^{DY} \over dE_T} \right] \label{sexp}
\ee
by normalizing the measured ratio at fixed $E_T$ by the measured integrated 
cross sections. The quantity (\ref{sexp}) can be directly computed 
in the Glauber theory formalism outlined above.  
Using the value of the pre-resonance absorption cross section (\ref{4}), 
determined from the analysis of $p-A$ data, we have found a good 
agreement with the $E_T$-integrated $O-Cu$, $O-U$ and $S-U$ data and with 
$E_T$ distributions measured in $S-U$ collisions, as we shall shortly show. 

For \P~production, the situation changes. The data for the integrated
and the differential survival probabilities are 
considerably lower than what nuclear absorption predicts, and the
additional suppression moreover increases with increasing $E_T$. 
We therefore need to include the effect of 
additional \P\ suppression on \J~production. 
The branching ratio for the reaction $\p \to \j$ is 0.57; therefore the 
$\psi/\psi'$ ratio measured in $pp$ and $pA$ collisions \cite{Carlos}
implies that $8\pm 2$ \% of the observed \J's are
due to \P~decay. Since the \P\ is suppressed in $S-U$ collisions,
the corresponding fraction of the observed \J's must be suppressed
as well. 
This correction reduces theoretical predictions on the average by 
$\simeq 5\%$. 
We show in Fig.\ 5 (left) the resulting corrected
theoretical $E_T$ dependence of \J~survival probabilities. 
The agreement between the data and
predictions is seen to be excellent. 

I wish to stress that we do not need to invoke 
any additional sources of direct \J\ suppression (apart from the nuclear 
absorption of pre-resonance charmonium state) to describe the data. 
On the other hand, the additional \P\ suppression 
found in $S-U$ collisions clearly indicates the presence of produced 
matter at the stage when charmonium states are formed. 
The agreement of our Glauber calculations with the measured \J\ 
survival probabilities shows, however, that this matter cannot break up 
\J\ states. This can be explained by the smallness of \J\ dissociation rate 
in confined hadronic gas \cite{KS}-\cite{KMS} advocated in sections 2 and 3.
\vskip0.2cm
\begin{figure}[h]
\noindent{\large{$S_{AB}^{J/\psi}$}} 
\begin{center}
\epsfxsize14cm
\mbox{\epsfig{file=qm96f8.ps,bbllx=2.5cm,bblly=8.8cm,bburx=21.5cm,bbury=17.cm,width=14.cm,angle=0}}
\end{center}
\hskip12cm {\large{$E_T$}}\\
\vskip0.3cm
\noindent Figure 5. $J/\psi$ suppression in $S-U$ (left) and $Pb-Pb$ 
(right) collisions; the NA38/50 data \cite{Carlos} (black points) 
are compared to the 
Glauber theory predictions \cite{KLNS} (grey points) 
with $\sigma_{abs}=6.3\pm 0.6$ mb.
\label{supp}
\end{figure}

\subsection{Pb-Pb Collisions}
\vskip0.3cm
We can further check our approach in $Pb-Pb$ collisions, since
the NA50 experiment \cite{MG} is equipped with 
a zero degree calorimeter (ZDC), which determines at each $E_T$ the
associated number of projectile spectators -- those projectile
nucleons which reach the ZDC with their full initial energy $E_{in}=158$
GeV/c. This additional information is important, since it uniquely
identifies the peripherality of the collision. 
Denoting the projectile as $A$, the number of projectile
spectators is evidently $A - N_w^A$, with $N_w^A$ of the $A$
nucleons in the projectile wounded. We thus have
$E_{ZDC}=(A-N_w^A)E_{in}$; using $\bar{E}_T=q~\bar{N}_w$, we predict 
\cite{KLNS} the
$E_T-E_{ZDC}$ correlation which agrees very well with the 
measured one \cite{MG}, \cite{web}.
\par
We are now ready to address the \J\ production in the NA50 experiment 
\cite{MG}.
The $Pb-Pb$ results, plotted as a function 
of the average path $L$ of \J\ in nuclear matter, clearly show strong 
additional suppression beyond the expected on the basis of 
$\sim exp(-\rho_0 \sigma_{abs} L)$ 
dependence \cite{MG}. The conclusion on the ``anomalous" nature of this 
suppression, however, crucially depends on the magnitude of $L$, assigned 
to the $Pb-Pb$ points. We would like therefore first to check the $L$ 
assignment of the NA50 Collaboration in our approach, which directly 
gives the \J\ survival probability at a given $E_T$. The results are 
presented in Fig.\ 5 (right). One can see that while the lowest $E_T$ point 
is still marginally consistent with Glauber theory, the suppression observed 
at higher $E_T$  
indeed goes significantly beyond expected. 
Equating our calculated survival probability to the form used by the NA50,
$S_{Gl}=exp(-\rho_0 \sigma_{abs} L)$, 
we confirm the NA50 $L$ assignments \cite{MG}. Since, as we have shown, 
the Glauber theory approach 
has been extremely successful in reproducing the bulk of \J\ production data 
in $p-A$ and $A-B$ collisions, the suppression observed in $Pb-Pb$
indeed can be called ``anomalous".

\section{IS THE QUARK-GLUON PLASMA DISCOVERED?}

\noindent Before we address this provocative question, 
posed at this Conference also 
by \mbox{J.-P. Blaizot} \cite{JP} and C.-Y. Wong \cite{Wong}, let us 
consider the possible differences between the collision dynamics in 
$S-U$ and $Pb-Pb$ systems. 
The success of Glauber theory in describing the \J\ suppression in 
$S-U$ collisions and its failure in $Pb-Pb$ points to a 
difference in the properties of matter produced in these two reactions.
We shall try to describe this difference in terms of two variables, 
one of which characterizes the energy density of produced matter, 
and the other its degree of equilibration. 

In Glauber theory, the initial energy density achieved in the 
collision is proportional to the density of wounded nucleons $n_w$ 
(see Eq.\ (\ref{9})) in the transverse plane. The average energy densities 
achieved in $S-U$ and $Pb-Pb$ collisions are almost identical; 
at first glance this suggests that the matter seen by produced \J's 
should be the same in both cases. This is not so, however, for two reasons. 
First, the profile of the energy density in the two systems is different: 
central $Pb-Pb$ collisions produce a ``hot core", inside which the energy 
density is higher than the highest one attainable in $S-U$ 
system by about $25\%$. Second, the \J's are produced mostly in this 
central region (see Eq.\ (\ref{6})), and thus feel the matter which 
is hotter than average. These two effects combined lead to significant 
difference in the energy densities of matter seen by \J's in $S-U$ and 
$Pb-Pb$ collisions. This is illustrated in Fig.\ 6, where we plot 
the ratio of experimental \J\ suppression to the Glauber theory 
predictions versus the average density of matter seen by \J\ (to calculate 
this latter quantity, we convolute the density distribution with the \J\ 
production profile). 
\begin{figure}[h!]
\hskip1cm {\large{${S_{exp}^{J/\psi} / S_{Gl}^{J/\psi}}$}}
\begin{center}
\epsfxsize9.5cm
\mbox{\epsfig{file=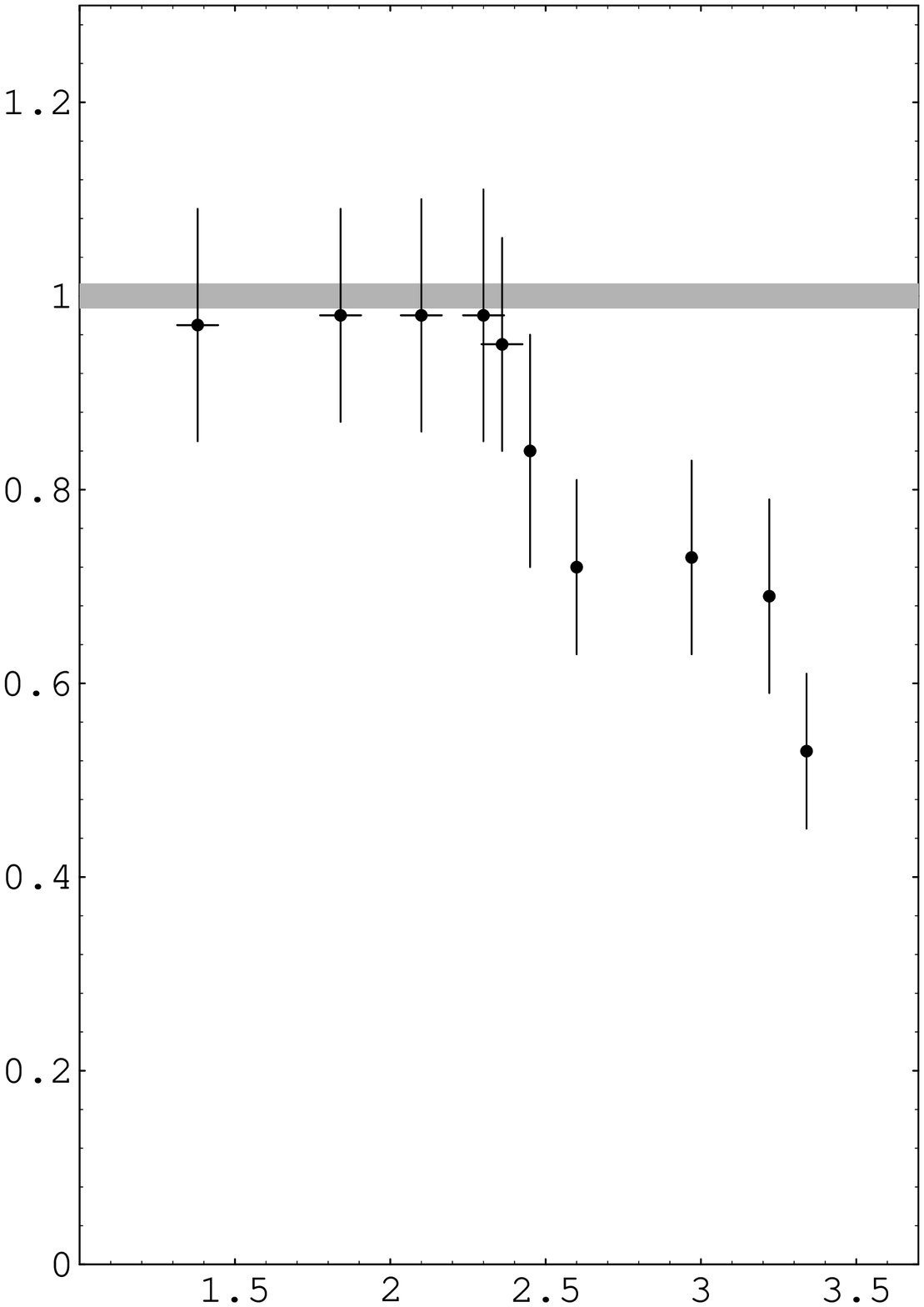,bbllx=2.5cm,bblly=2.cm,bburx=21.5cm,bbury=25.cm,width=9.5cm,angle=0}}
\end{center}
\hskip10cm {\large{$\bar{n}_w^{J/\psi}, {\rm fm}^{-2}$}}\\
\vskip0.3cm
\noindent Figure 6. The ratio of experimental $J/\psi$ suppression in $S-U$ 
\cite{Carlos} (crosses) and $Pb-Pb$ \cite{MG}  
(circles) collisions to the Glauber theory predictions versus the average 
density of wounded nucleons seen by $J/\psi$ (the latter quantity is 
proportional to the energy density).
\label{ratio}
\end{figure}
Apart from being more dense, the matter seen by \J\ is also likely to be 
more thermalized. We can quantify this statement introducing the variable 
\be
\kappa = {\nu + 1 \over w}, \label{kappa}
\ee
where $\nu$ is the number of inelastic $NN$ collisions and 
$w$ is the number of wounded 
nucleons these collisions produce. The value of $\kappa$ 
tells how many times, on the average, each wounded nucleon was hit. 
In a $pp$ collision, 
$\nu=1$ and $w=2$, so that $\kappa=1$. In $pA$ collisions, the number 
of wounded nucleons in the target is equal to the number of collisions 
\cite{Bialas}, 
which, after taking into account the wounded projectile nucleon, again 
yields $\kappa=1$. In nucleus-nucleus collisions, however, the value of 
$\kappa$ can exceed unity because the nucleons once wounded can collide 
again and again. These collisions can break down the independence of 
fragmentation of wounded nucleons  
and provide initial conditions for the onset of collective behaviour in the 
system. Indeed, when $\kappa>1$, the partons from different wounded 
nucleons interact, which is a necessary initial stage for producing 
deconfined matter -- at large $\kappa$, partons can no longer be 
attributed to a particular pair of wounded nucleons and overlap 
in the transverse plane. It is evident that in nuclear collisions 
$\kappa$ grows with atomic number and/or energy 
(since the number of collisions depends on the inelastic cross section).
In central $S-U$ collisions at SPS energy, the Glauber theory calculation 
yields $\bar{\kappa}_{SU}\simeq 1.7$, whereas for a central $Pb-Pb$ 
collision we find $\bar{\kappa}_{PbPb}\simeq 2.4$. The $Pb-Pb$ value 
thus is as far from the $S-U$ one as the $S-U$ is from $pA$.  
Central $Pb-Pb$ collisions therefore are likely to produce not only more 
dense, but also more thermalized matter.

It is important to note that the value of $\kappa$ and its variation with 
centrality can be determined in a model--independent way directly 
from the experimental data. Indeed, the number 
of collisions $\nu$ is proportional to the number of produced Drell-Yan 
pairs, and the number of wounded nucleons to the produced 
transverse energy 
(see Eq.\ (\ref{propw})), 
so at a given $E_T$ one has
\be
\kappa (E_T) \sim q {N_{DY}(E_T) \over E_T} = {q \over E_T}\  
{1 \over \sigma_{DY}\Delta E}\ \int_{\Delta E} {d\sigma_{DY} \over dE_T} 
dE_T, \label{kapexp}
\ee
where $N_{DY}(E_T)$ is the number of Drell-Yan pairs associated 
with a given $E_T$ bin of the width $\Delta E_T$.
 
We thus conclude that the matter seen by \J\ in central $Pb-Pb$ collisions 
is more dense and more thermalized, so the occurence of new phenomena 
at least cannot be excluded {\it a priori}. 
If the observed \J\ suppression were indeed to be interpreted as a 
signal of deconfinement phase transition, apart from being ``anomalous", 
it also has to exhibit a threshold behaviour. This feature seems to be 
present in the data (see Fig.\ 6): a slight increase in the energy 
density induces dramatic deviation from the trend established previously. 
Moreover, if one considers the profile of the density $n_w$ and $\kappa$ in 
the transverse plane and assumes that {\it all} \J's produced in the region 
where $n_w > n_{SU}^{max},\ \kappa > \kappa_{SU}^{max}$ are dissociated, 
the resulting suppression falls below the data points by only $10-15\%$ 
(similar analysis has been presented at this Conference by J.-P. Blaizot 
\cite{JP}).
This shows that 
the observed suppression is almost as strong as we could possibly accomodate! 

Can we still find a ``conventional" explanation of the NA50 effect? 
It is of course too early to try to answer this question; let us therefore 
limit ourselves to some preliminary observations.
Since Glauber calculations show that nuclear absorption of pre-resonance 
state cannot explain the $Pb-Pb$ data, a conventional 
explanation has to invoke 
additional suppression of \J's in the produced confined matter. 
This suppression would be characterized by a smooth (basically, exponential) 
dependence on the density of produced matter. Looking at Fig.\ 6, one 
realizes that it would not be easy to fit the data by this kind of 
a smooth dependence -- the fits would most likely overestimate the slope of 
the $S-U$ data and underestimate it for the $Pb-Pb$ points. 
The physical reason for this is transparent -- if hadronic 
comovers 
do not induce an additional \J\ suppression in $S-U$ collisions, it is 
difficult to make 
them effective in $Pb-Pb$ system. 
One may try 
to assume a larger density variation with $E_T$ (sometimes the  
density is assumed to be directly proportional to $E_T$ \cite{Sean}). 
Indeed, a large variation 
of density is possible in very central collisions due to fluctuations 
in the number of produced hadrons. However the measured 
$E_T-E_{ZDC}$ correlation discussed in section 4.3 shows that 
in the presented $Pb-Pb$ data the variation of $E_T$ results from the 
variation 
of the collision centrality. Even the highest $E_T$ point of NA50 is 
not entirely in the fluctuation domain, and corresponds to the mean 
impact parameter of $\bar{b}\simeq 2$ fm. In this regime, the  
$E_T$ measured by the NA50 varies by more than four times, but this leads 
only to $\sim 30\%$ variation of the initial energy density (see Figs.\ 5 
and 6). Additional constraint on a conventional scenario is imposed by 
the $\psi'/\psi$ ratio: 
if the density of comovers in $Pb-Pb$ were much higher than the density 
in $S-U$, this would imply a smaller $\psi / \psi'$ ratio in 
the former case -- 
the prediction that would bring us in conflict with the data \cite{MG}.      
A possible way to describe the $Pb-Pb$ points conventionally would be to 
decrease artificially the value of 
the nuclear absorption cross section $\sigma_{abs}$, leaving thus room 
for comover absorption already in $S-U$ collisions (the need  
for comover effects to explain the $S-U$ data was advocated at this 
Conference by S. Gavin \cite{Sean}). This, however, 
would contradict to the $pA$ data, that fix the value of $\sigma_{abs}$ 
rather precisely (see section 4.1).
Of course, it remains to be seen if a convincing conventional explanation 
can eventually be found.

To summarize, the NA50 $Pb-Pb$ results indeed seem to suggest that a new 
mechanism of \J\ suppression sets in at higher energy densities. 
The observed effect can be considered as a strong evidence of   
some kind of deconfinement in nuclear collisions. What can we do 
to turn this evidence into a proof, or to discard it?

\section{WHAT HAS TO BE DONE NEXT?}

New precision data on quarkonium production coming from 
CERN SPS, Fermilab, HERA and elsewhere allow us today to get rid of 
many uncertainties inherent to the analyses of \J\ suppression 
over the years. 
A coherent picture, providing a good description of 
the bulk of existing $pp$, $pA$ and $AB$ data, has started to emerge -- 
this makes us 
ready to recognize and study unusual phenomena.
It is therefore particularly important to learn more about  
 the onset of anomalous behaviour of \J\ suppression 
seen by the NA50 Collaboration. More statistics and more data points, 
both in the transition regime of small $E_T$ and in the fluctuation 
region of the highest $E_T$, are needed to establish the threshold 
behaviour suggested by the present data. 
An important information would be also provided by the \J\ transverse 
momentum  distributions \cite{PT}.

Direct measurement of the low-energy \J\ absorption cross section  
 in the proposed inverse kinematics experiment \cite{KS},\cite{KS4}  
has become possible 
with the advent of $Pb$ beam at CERN SPS. This experiment would 
allow us to directly constrain the \J\ absorption possible in a 
hadronic medium, 
providing important additional check of the deconfinement transition as 
the cause of ``anomalous" \J\ suppression.     

Heavy quarkonium represents a rare example of a strongly interacting 
system that is simple enough to be 
systematically analyzed by the current theoretical methods. 
It has already proved to be extremely useful for understanding 
the properties of QCD and its ground state - the vacuum. 
I believe that quarkonium will tell us much also about the critical 
behaviour of QCD matter produced by relativistic heavy ion collisions.

\vskip0.3cm
{\bf Acknowledgements}
\vskip0.2cm
\noindent The results presented here were obtained together with 
 H. Satz; I wish to thank him for an enjoyable collaboration. 
I am grateful to 
L. McLerran for his collaboration, valuable comments and encouragement. 
Very important contributions to the work presented here have been made by  
C. Louren\c{c}o, M. Nardi, A. Syamtomov, X.-N. Wang, X.-M. Xu and 
G.M. Zinoviev. Stimulating and 
enlightening discussions of various aspects of this work with   
J.-P. Blaizot, E. Braaten, A. Capella, J. Ellis, K.J. Eskola, 
S. Gavin, M. Ga\'{z}dzicki, \mbox{C. Gerschel,} \mbox{M. Gonin,} 
M. Gyulassy, M. Jacob,  A.B. Kaidalov, 
F. Karsch, L. Kluberg, \mbox{T. Matsui,} \mbox{A.H. Mueller,} B. M\"{u}ller, 
J.-Y. Ollitrault, J. Schukraft, G. Schuler, \mbox{R. Stock,} 
\mbox{A.I. Vainshtein} and R. Vogt are gratefully acknowledged.


\begin{thebibliography}{999}

\bibitem{MS}
{T. Matsui and H. Satz, \PL B 178 (1986) 416.}
\bibitem{NA38}
{The NA38 Collaboration, C. Baglin et al., \PL B 220 (1989) 471; B 251 (1990) 
 465, 472; B225 (1991) 459.}
\bibitem{Blaizot}
{for a review, see J.-P. Blaizot and J.-Y. Ollitrault, in: 
``Quark-Gluon Plasma", \\ 
R.C. Hwa (Ed.), World Scientific, Singapore, 1990, p.631.}
\bibitem{Carlos}{C. Louren\c{c}o, {\it in these Proceedings}}.
\bibitem{Pat}{P. McGaughey, {\it in these Proceedings}}.
\bibitem{San}{A. Sansoni, {\it in these Proceedings}}.
\bibitem{Braa}{E. Braaten, {\it in these Proceedings}}.
\bibitem{Peskin}
{M. E. Peskin, \NP B 156 (1979) 365;\\
G. Bhanot and M. E. Peskin, \NP B 156 (1979) 391.}
\bibitem{Kaidalov}{A. Kaidalov, in {\sl QCD and High Energy
Hadronic Interactions}, J. Tr\^an Thanh V\^an (Ed.), Editions Frontieres,
Gif-sur-Yvette, 1993, p.601.}
\bibitem{KS}
{D. Kharzeev and H. Satz, \PL B 334 (1994) 155.}
\bibitem{KS1}
{D. Kharzeev and H. Satz, in: ``Quark-Gluon Plasma 2", R.C. Hwa (Ed.), 
World Scientific, 1995, p.395.}
\bibitem{KMS}
{D. Kharzeev, L. McLerran and H. Satz, \PL B 356 (1995) 349.}
\bibitem{S1}
{H. Satz, Nucl. Phys. A590 (1995) 63c.}
\bibitem{K1}
{D. Kharzeev, CERN-TH/95-342, nucl-th/9601029.}
\bibitem{KS2}
{D. Kharzeev and H. Satz, Phys. Lett. B 366 (1996) 316.}
\bibitem{GH}
{A. Capella, J.A. Casado, C. Pajares, A.V. Ramallo and J. Tr\^an Thanh V\^an, 
\PL B 206 (1988) 354;\\
C. Gerschel and J. H\"ufner, \PL B 207 (1988) 253; Zeit.Phys. C 56 (1992) 
391.}
\bibitem{MG}
{M. Gonin, {\it in these Proceedings.}}
\bibitem{KS4}
{D. Kharzeev and H. Satz, \PL B 356 (1995) 365.}
\bibitem{Parisi}
{G. Parisi, \PL B 43 (1973) 207; B 50 (1974) 367.}
\bibitem{KSSZ}
{D. Kharzeev, H. Satz, A. Syamtomov and G. Zinoviev, CERN-TH/96-72, and\\  
{\it in preparation.}}
\bibitem{JE}
{J. Ellis, \NP B 22 (1970) 478;\\
R.J. Crewther, \PL B 33 (1970) 305; \PRL 28 (1972) 1421;\\
M.S. Chanowitz and J. Ellis, \PL B 40 (1972) 397.}
\bibitem{EK}
{J. Ellis and D. Kharzeev, CERN-TH/96-151.}
\bibitem{VZ}
{M.B. Voloshin and V.I. Zakharov, \PRL 45 (1980) 688;\\
A.B. Kaidalov and P.E. Volkovitsky, \PRL 69 (1992) 3155;\\
M. Luke, A.V. Manohar and M.J. Savage, \PL B 288 (1992) 355.}
\bibitem{CS}
{C.H. Chang, \NP B 172 (1980) 425;\\
E.L. Berger and D. Jones, \PR D 23 (1981) 1521;\\
R. Baier and R. R\"uckl, \PL B 102 (1981) 364; \ZP C 19 (1983) 251.}
\bibitem{Ger}
{G.A. Schuler, CERN-TH.7170/94; Phys. Rep., {\it to appear.}}
\bibitem{Gav} 
{R. Gavai, D. Kharzeev, H. Satz, G. Schuler, K. Sridhar and R. Vogt, 
Int. J. Mod. Phys. A 10 (1995) 3043.}
\bibitem{SVZ} 
{M.A. Shifman, A.I. Vainshtein and V.I. Zakharov, \NP B 147 (1979) 385, 448.} 
\bibitem{Vol}
{M.B. Voloshin, \NP B 154 (1979) 365; Sov.J.Nucl.Phys. 36 (1982) 143.}
\bibitem{NR}
{G.T. Bodwin, E. Braaten and G.P. Lepage, \PR D 51 (1995) 1125.}
\bibitem{Mue}
{A.H. Mueller, Nucl. Phys. B 415 (1994) 373.}
\bibitem{KMeS}
{F. Karsch, M. T. Mehr and H. Satz, \ZP C 37 (1988) 617.}
\bibitem{XW}
{X.-M. Xu, D. Kharzeev, H. Satz and X.-N. Wang, Phys. Rev. C 53 (1996) 3051.}
\bibitem{KLNS}
{D. Kharzeev, C. Louren\c{c}o, M. Nardi and H. Satz, {\it in preparation}.}
\bibitem{deJager}
{C.W.deJager, H.deVries and C.deVries, At.Data and Nucl.Data Tables 
14(1974)485.}
\bibitem{Bialas}
{A. Bialas, M. Bleszy\'{n}ski and W. Czy\.{z}, Nucl. Phys. B111 (1976) 461.}
\bibitem{GV}
{S. Gavin and R. Vogt, Nucl. Phys. B345 (1990) 104.}
\bibitem{Stock}
{The NA49 Collaboration, T. Alber et al., Phys. Rev. Lett. 75 (1995) 3814;\\
The NA35 Collaboration, J. B\"{a}chler et al., Z. Phys. C52 (1991) 239.}
\bibitem{web}
{The NA50 Collaboration WWW page: http://www.cern.ch/NA50/figures.html}
\bibitem{JP}
{J.-P. Blaizot, {\it in these Proceedings}.}
\bibitem{Wong}
{C.-Y. Wong, {\it in these Proceedings}}.
\bibitem{Sean}
{S. Gavin, {\it in these Proceedings}}.
\bibitem{PT}
{F. Karsch and R. Petronzio, Phys. Lett. B 193 (1987) 105;\\
J.-P. Blaizot and J.-Y. Ollitrault, Phys. Lett. B 199 (1987) 499.}

\end{thebibliography}
\end{document}